\documentclass[osajnl,twocolumn,showpacs,superscriptaddress,10pt]{revtex4-1}
\usepackage{amsmath,amssymb,graphicx}

\usepackage{mathtools}
\usepackage{xfrac}

\begin{document}

\title{Thermally Controlled Comb Generation and Soliton Modelocking in Microresonators}

\author{Chaitanya Joshi}
\email{Corresponding author: chaitanya.joshi@columbia.edu}
\affiliation{Department of Applied Physics and Applied Mathematics, Columbia University, New York, NY 10027, USA}
\affiliation{School of Applied and Engineering Physics, Cornell University, Ithaca, NY 14853, USA}

\author{Jae K. Jang}
\affiliation{Department of Applied Physics and Applied Mathematics, Columbia University, New York, NY 10027, USA}

\author{Kevin Luke}
\affiliation{School of Electrical and Computer Engineering, Cornell University, Ithaca, NY 14853, USA}

\author{Xingchen Ji}
\affiliation{School of Electrical and Computer Engineering, Cornell University, Ithaca, NY 14853, USA}
\affiliation{Department of Electrical Engineering, Columbia University, New York, NY 10027, USA}

\author{Steven A. Miller}
\affiliation{School of Electrical and Computer Engineering, Cornell University, Ithaca, NY 14853, USA}
\affiliation{Department of Electrical Engineering, Columbia University, New York, NY 10027, USA}

\author{Alexander Klenner}
\affiliation{Department of Applied Physics and Applied Mathematics, Columbia University, New York, NY 10027, USA}

\author{Yoshitomo Okawachi}
\affiliation{Department of Applied Physics and Applied Mathematics, Columbia University, New York, NY 10027, USA}

\author{Michal Lipson}
\affiliation{Department of Electrical Engineering, Columbia University, New York, NY 10027, USA}

\author{Alexander L. Gaeta}
\affiliation{Department of Applied Physics and Applied Mathematics, Columbia University, New York, NY 10027, USA}

\begin{abstract}We report the first demonstration of thermally controlled soliton modelocked frequency comb generation in microresonators. By controlling the electric current through heaters integrated with silicon nitride microresonators, we demonstrate a systematic and repeatable pathway to single- and multi-soliton modelocked states without adjusting the pump laser wavelength. Such an approach could greatly simplify the generation of modelocked frequency combs and facilitate applications such as chip-based dual-comb spectroscopy.
\end{abstract}


\maketitle 

Optical frequency comb generation is a revolutionary technology that enables new capabilities in spectroscopy \cite{Diddams2007}, time and frequency metrology \cite{Udem2002},  optical arbitrary waveform generation \cite{Jiang2007}, low-noise radio frequency (RF) signal generation \cite{Fortier2011}, and optical clockwork \cite{DiddamsClockwork,Newbury2011}. Recently, there has been a significant development in frequency comb technology based on microresonators with demonstrations in calcium fluoride \cite{Savchenkov2008}, magnesium fluoride (MgF$_2$) \cite{Liang11,Herr2014}, silica \cite{DelHaye2007,Jiang2012,vahala}, aluminum nitride \cite{Jung14}, diamond \cite{Haussmann2014}, silicon \cite{Griffith2015}, and silicon nitride (Si$_3$N$_4$) \cite{Foster11,Wang13,Brasch2016,kasturi,huang2015,Xue2015}. Si$_3$N$_4$ has emerged as a particularly attractive platform for chip-scale frequency comb generation, since it uses a CMOS-compatible fabrication process \cite{Moss2013} and allows for integration of electronics and optical elements in a compact, robust, and portable device. This will open up applications of frequency combs to a wider range of environments as compared to current frequency comb sources that are mostly limited to controlled laboratory environments.  
 
In order to utilize microresonator-based combs for precision time and frequency applications, the comb output must be in the low-noise modelocked state \cite{kasturi,Herr2014}. In microresonators, the generation of a single- or multi-soliton state in the ring corresponds to passive modelocking and ultrashort pulse formation. To date, single-soliton states have been generated in microresonators using pump frequency tuning in MgF$_2$ \cite{Herr2014}, silica \cite{vahala}, and Si$_3$N$_4$  \cite{kasturi}, and with pump power control in Si$_3$N$_4$   \cite{Brasch2016}. A theoretical study of a definitive route to soliton modelocking in microresonators by varying the frequency or the power of the pump laser has been previously described \cite{Lamont13,Matsko11,Coen13,Chembo2013,Villegas15}. By tuning the frequency of the pump laser, the generated comb transitions through a sequence of distinct states to ultimately reach the low-noise soliton state \cite{Luo15}. 

However, the use of laser frequency tuning in comb generation has drawbacks. The performance of the comb is limited by the linewidth and the amplitude noise of the pump since the dynamics of frequency comb generation is governed by parametric four-wave mixing (FWM) \cite{Herr2012,fwmphase}. Tunable lasers suffer from the drawback that they are relatively noisy and have a broader linewidth that is usually of the order of a hundred kHz. In contrast, fixed-frequency lasers can be operated with significantly lower noise and narrower linewidths than tunable lasers as the laser cavity is monolithic, and the lack of moving components eliminates sources of noise. Additionally, by locking the output to a frequency reference, the linewidth can be further reduced. Recent demonstrations of locked fixed-frequency lasers have shown linewidths of $\leqslant$40~mHz \cite{Kessler2012}. Using a locked low-noise and narrow linewidth fixed-frequency laser as the pump in place of tunable lasers will significantly reduce the noise on the generated comb lines. In addition, with the pump laser frequency fixed, the only uncertain parameter to fully determine the frequencies of the comb lines becomes the free spectral range (FSR). Locking the FSR will allow for a fully-stabilized comb where the frequency of each comb line can be determined. Furthermore, control of the cavity resonance frequency rather than the pump frequency allows for the simultaneous generation of modelocked frequency combs in multiple resonators on a single chip using a single fixed-frequency pump laser. This is essential for applications such as dual-comb spectroscopy \cite{Keilmann04} that requires two frequency comb sources that have a slightly different FSR. Thermal tuning of the resonance has previously been demonstrated using pump power control \cite{Brasch2016}, electro-optic tuning \cite{Jung14},and integrated heaters \cite{Xue2015} for comb generation. 

Here, we report the first demonstration of soliton modelocking in Si$_3$N$_4$ microresonators using integrated heaters for thermal control of the cavity resonance. Current control of integrated heaters results in a change in the waveguide refractive index due to the thermo-optic effect,  which changes the resonant frequency of the cavity \cite{Cunningham10}. We present a repeatable and systematic method for achieving low-noise single- and multi-soliton states using a narrow linewidth fixed-frequency laser as the pump.

\begin{figure}[htbp]
\centering
\includegraphics[width=\linewidth]{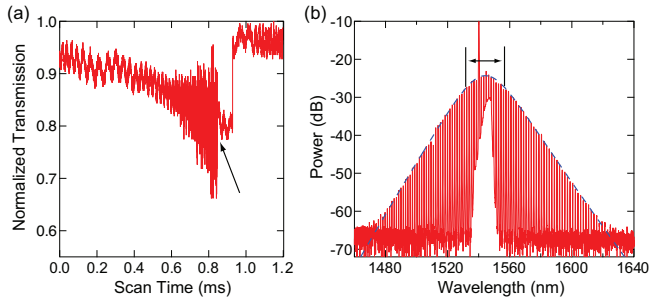}
\caption{(a) The pump power transmission as the tunable laser frequency is scanned across the resonance. The step-like structure characteristic of soliton formation is indicated by the arrow. (b) The measured optical spectrum for a single-soliton modelocked state with the fitted sech$^2$-pulse spectrum (blue dashed line). The 3 dB bandwidth of the soliton is 24 nm.}
\label{fig:piezo}
\end{figure}

In our experiment, we use an oxide-clad Si$_3$N$_4$ microring resonator with a  FSR of 200 GHz and a cross section of 950~$\times$~1500~nm. The waveguide cross section is chosen such that a region of anomalous group-velocity dispersion (GVD) exists near the pump wavelength \cite{Okawachi14}. Initially, we characterize the resonator using a tunable laser at 1540 nm. We amplify the laser using an erbium-doped fiber amplifier (EDFA) and couple 56 mW of power into the bus waveguide for comb generation. We change the detuning of the laser with respect to the resonance frequency of the microresonator by scanning the laser frequency using piezoelectric tuning. We monitor the transmitted power at the pump mode using a fast photodiode ($\geqslant$12.5 GHz) and observe the optical spectrum on an optical spectrum analyzer (OSA). We modulate the laser frequency using a triangular waveform and record the pump transmission over one resonance scan as seen in Fig. \ref{fig:piezo}(a). We observe the characteristic step-like structure indicative of the transition into modelocked soliton states as demonstrated in previous work \cite{Herr2014}. Furthermore, the measured optical spectrum is in agreement with the fitted sech$^2$ spectrum denoted by the dashed blue curve in Fig. \ref{fig:piezo}(b). 

\begin{figure}[htbp]
\centering
\includegraphics[width=\linewidth]{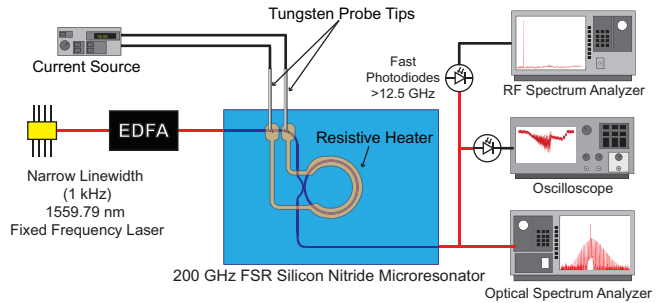}
\caption{Experimental setup for generation and characterization of soliton modelocked states in Si$_3$N$_4$ microresonators. We characterize the optical spectrum, RF amplitude noise spectrum, and transmitted pump power simultaneously. Integrated resistive heaters are used to tune the resonance frequency to generate frequency combs.}
\label{fig:setup}
\end{figure}

To demonstrate the feasibility of generating soliton modelocked combs using thermal tuning, we use a continuous-wave fixed-frequency laser with a narrow linewidth of 1 kHz at a wavelength of 1559.79 nm as the pump laser. We amplify the output using a high power EDFA and couple 71 mW into the bus waveguide using a lensed fiber. For comb generation, the nearest resonance frequency is tuned by varying electric current through the integrated platinum resistive heaters, which have an electrical resistance of 240 $\Omega$. We require about 150 mW of electrical power to tune the nearest resonance frequency close to the pump laser frequency. We monitor the optical spectrum, RF spectrum, and transmitted pump power of the generated comb. Figure \ref{fig:setup} shows the setup used for generation and characterization of frequency combs using thermal tuning. The free-space output is collected using a combination of an aspheric lens and a collimator and coupled into a fiber. The light is then split 80:20 using a fiber power splitter, and the smaller fraction of the power is sent to the OSA to record the generated comb spectrum as it transitions through the comb formation dynamics. The remaining power is sent to a wavelength division multiplexing (WDM) filter with a 100-GHz transmission window centred at the pump wavelength. The transmitted light through the filter is sent to a fast photodiode ($\geqslant$12.5~GHz) to monitor the pump transmission as the resonance is tuned. The reflected light from the WDM filter is sent to a second fast photodide that is used to monitor the RF amplitude noise on the generated comb close to DC (0-900~MHz) using a RF spectrum analyzer at a resolution bandwidth of 100~kHz.

\begin{figure}[htbp]
\centering
\includegraphics[width=\linewidth]{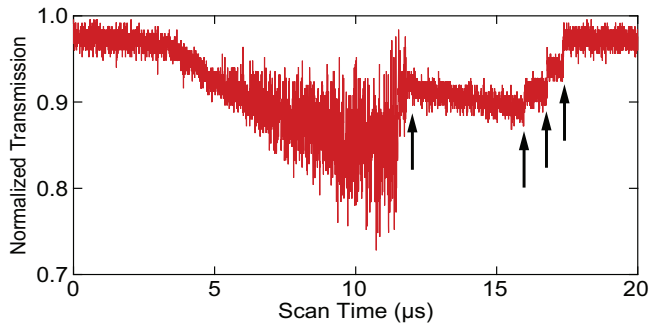}
\caption{Oscilloscope trace of the pump transmission as the current on the integrated heater is modulated with a triangular waveform. The steps indicated by the arrows are characteristic of transitions between different multi-soliton states.}
\label{fig:trace}
\end{figure}

We apply a triangular modulation to the heater current to scan the cavity resonance near the laser wavelength. This corresponds to a 5 mW change in the electrical power applied to the heater. Similar to the case with laser frequency tuning, we observe the characteristic steps in the pump power transmission (Fig.~\ref{fig:trace}), in which each step is indicative of a transition from a higher to a lower number of solitons. The final step consists of a transition from the single-soliton state to the laser frequency dropping out of the cavity resonance.  

\begin{figure}[htbp]
\centering
\includegraphics[width=\linewidth]{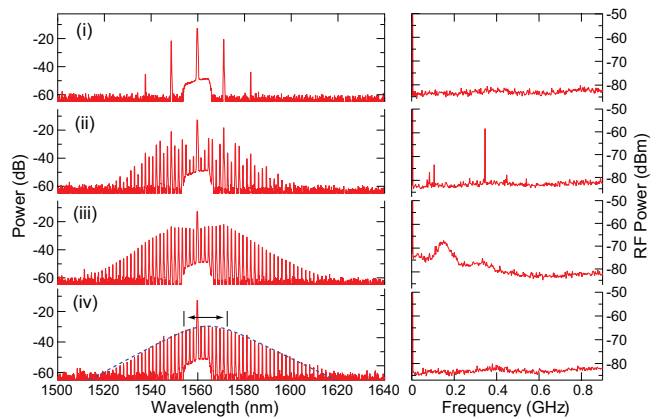}
\caption{Evolution of the generated frequency comb spectrum as the cavity resonance is tuned by varying the heater current. The optical and RF spectra as the comb evolves correspond to (i) the initial cascaded FWM, (ii) the mini-comb formation, (iii) the broadband high-noise regime, with the plateau-like optical spectrum and broad noise peak and (iv) the low-noise single-soliton state with a fitted sech$^2$-spectral profile (blue dashed curve). The 3 dB bandwidth of the soliton is 20 nm.}
\label{fig:evol}
\end{figure}

We study the evolution of the comb generation process and observe transitions into various comb states as we change the resonance frequency with respect to the laser frequency (Fig.~\ref{fig:evol}). As the power in the resonator builds up, we see the primary sidebands form at the parametric gain peak due to degenerate FWM [Fig.~\ref{fig:evol}(i)]. The RF amplitude noise at this stage is low since it corresponds to parametric oscillation for a single signal and idler pair. Tuning the resonance further, we see mini-comb formation [Fig.~\ref{fig:evol}(ii)] with natively spaced lines near the primary sidebands. The interaction of the separate mini-combs within the cavity manifests on the RF spectrum as a sharp spike. Subsequently, we observe the transition into the broadband high-noise regime [Fig.~\ref{fig:evol}(iii)], and the RF noise peak also broadens. Finally, the system undergoes a transition to the single-soliton state [Fig.~\ref{fig:evol}(iv)] with the reduction of the RF noise and the optical spectrum showing the characteristic shape of a sech$^2$-pulse spectrum.

It is important to note that the single-soliton state can be achieved by scanning through the cavity resonance using thermal tuning at a sufficiently high speed. The temperature of the ring depends on the coupled pump power and the thermal time constant of the ring. The speed of the thermal scan affects the rate at which the coupled pump power in the ring changes. Thus the speed of the scan affects the temperature variation in the ring. The soliton state occurs at a certain equilibrium temperature inside the ring, and when we scan the resonance frequency at a slow rate (200~Hz) using a triangular modulation, we observe that the steps corresponding to the soliton formation are narrow and are not consistent from one scan to the next. Here, the scan is significantly slower than the thermal time constant,  and the temperature of the ring rises above the equilibrium soliton temperature, which prevents the system from reaching the soliton state consistently. At higher scan speeds (e.g.,~10~kHz), we see the steps on the pump transmission are wider and consistent from one scan to the next. Here, the thermal scan speed is closer to the thermal time constant of the ring, and the corresponding variation in temperature of the ring is smaller. The system is consistently able to reach the equilibrium temperature and the soliton state. Similar behavior has been previously reported in \cite{Herr2014} where the speed of the pump frequency scan affects the reproducibility of the soliton states.

\begin{figure}[htbp]
\centering
\includegraphics[width=\linewidth]{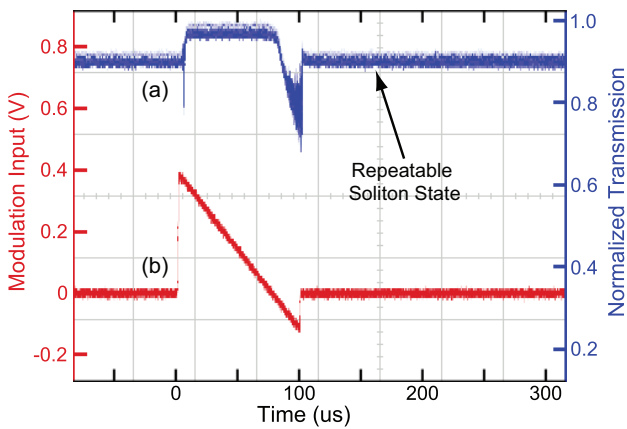}
\caption{(a) A persistence trace of the pump transmission recorded over 3 seconds. We see that the comb returns to the same soliton state over 15 consecutive traces. The modulation signal sent to the current source is shown in (b). The downward slope corresponds to a blue-shift of the resonance. The abrupt increase in current red-shifts the resonance and leads to the repeatable generation of the soliton state.}
\label{fig:control}
\end{figure}

We start the scan with the laser frequency blue detuned with respect to the resonance frequency of the ring and apply a downward ramp that is at a speed that enables the formation of the soliton state as explained above. This ramp blue shifts the resonance and the comb evolves as shown in Fig. \ref{fig:evol}. At the end of the ramp before terminating the scan, we apply a small rise in the current that corresponds to a red shift of the resonance. We repeat this current modulation every 200 ms and record a persistence trace of the transmitted pump power lasting 3 seconds. The transmission trace indicates clearly that the system reaches the same final soliton state over all 15 scans as seen in Fig. \ref{fig:control}(a). The tuning curve of the current modulation can be seen in Fig. \ref{fig:control}(b). We observe that the red detuning prior to terminating the scan makes the formation of the soliton state more repeatable as compared to a ramp signal without the red shift. The repeatability of the soliton state is affected by drift of the input fiber coupling that leads to fluctuations in the coupled power. In a packaged device, the issue of input coupling fluctuations will be eliminated since the pump laser will not physically drift with respect to the bus waveguide. A similar result was recently demonstrated using pump frequency tuning with the 'backward tuning' method that allows for repeatable soliton formation \cite{Karpov2016}. Our tuning curve for the resonance frequency [Fig. \ref{fig:control}(b)] is analogous to the 'backward tuning' method presented in that work with a blue shift using the downward slope and a subsequent red shift due to the upward ramp that allows repeatable soliton formation. Furthermore, we can also switch from a higher number of solitons to a lower number of solitons by slowly increasing the heater current and red shifting the resonance once it is in a stable multi-soliton state. 

\begin{figure}[htbp]
\centering
\includegraphics[width=\linewidth]{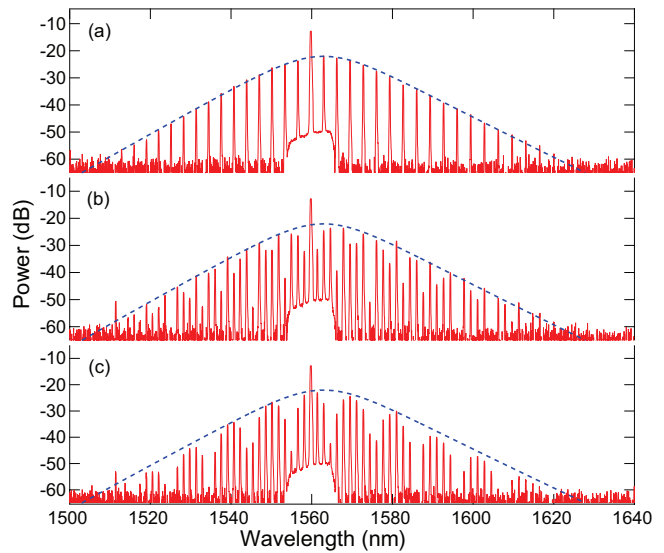}
\caption{Measured spectra for three different multi-soliton states (a), (b), and (c). The modulations on the spectra are due to the spectral interference among multiple modelocked solitons within one roundtrip of the cavity (blue dashed line indicates a fitted sech$^2$ envelope for a single-soliton). The spectrum in (a) is indicative of a two soliton state with the pulses half a round trip apart.}
\label{fig:multisol}
\end{figure}

We can choose the final state of the frequency comb by adjusting the red shift before we terminate the scan. By modifying this termination point, we observe different multi-soliton states. The relative positions of the multiple solitons in the ring result in modulations on the sech$^2$-spectral profile. The measured multi-soliton spectra are depicted in Fig. \ref{fig:multisol}. Of particular interest is the spectrum depicted in Fig. \ref{fig:multisol}(a) where every other comb line in the spectrum is extinguished. This is indicative of a two-soliton state with the two solitons exactly half a roundtrip apart, corresponding to harmonic modelocking \cite{Hirano1969}.

In conclusion, we report the first demonstration of low-noise single-soliton states in Si$_3$N$_4$ microring resonators using thermal control of integrated heaters. We demonstrate a systematic and repeatable pathway to tune into single- and multi-soliton states by changing the electrical power on the heaters by 5 mW from 150 mW. The system also allows for progressive switching from a higher number of solitons in the cavity toward a single-soliton state. Thermal control enables the use of low-noise fixed-frequency lasers which will lead to monolithic design of a fully integrated chip-scale comb source. Furthermore, the technique will enable the simultaneous generation of multiple modelocked combs from a single pump source, which is critical for the realization of coherent spectroscopic applications such as dual-comb spectroscopy.
\newline \newline
\noindent \textbf{Funding.} 
Air Force Office of Scientific Research (FA9550-15-1-0303); Defense Advanced Research Projects Agency (W31P4Q-15-1-0015); A.K. acknowledges a postdoc fellowship from the Swiss National Science Foundation (P2EZP2\_162288)
\newline \newline
\noindent \textbf{Acknowledgements.} This work was performed in part at the Cornell Nano-Scale Facility, a member of the National Nanotechnology Infrastructure Network, which is supported by the NSF. We thank J. Ye and B. Bjork for useful discussion.



\end{document}